\documentclass[10pt]{article}
\usepackage{latexsym,amssymb,amsfonts,amsmath}
\usepackage{graphicx}

\begin{document}
\title{On solutions for the $b$-family of peakon equations}
 \author{{Zhaqilao${{\thanks {
 Corresponding author: Tel. : +86 471 4392483 : fax: +86 471 7383390. E-mail address:
zhaqilao@imnu.edu.cn}}}$}
\\
{\scriptsize{College of Mathematics Science, Inner Mongolia Normal
University, Huhhot 010022, People's Republic of China}}}

\date{}
\maketitle

%E-mails: zhaqilao@imnu.edu.cn and 1416528264@qq.com (Yuanli  Li)

\baselineskip 18pt  {\bf \large{Abstract:}} We investigate a family of peakon equations, labelled by two parameters $b$ and $\kappa$, all of which admit one-peakon solutions in a unified form. The well known Camassa-Holm equation and Degasperis-Procesi equation are derived from the b-family peakon equations by choosing $b=2$ and 3, respectively. For all values of $b$, their two peakon-type solutions are shown in difference form and in weak sense. At the same time, the dynamic behaviors are shown in difference phenomenons, such as the peaked waves collide elastically in the case $b=2$, while the peaked waves collide inelastically in other cases.

\vskip 1mm



\noindent \underline{\hspace*{14.8cm}}

\parskip 10pt
\setcounter{equation}{0}
\section{Introduction}
In recent years, there has been considerable interest in the following a family of partial differential equations \cite{0,1}
\begin{equation}\label{1}
u_t+b\kappa u_x-u_{xxt}+(b+1)uu_x
=bu_xu_{xx}+uu_{xxx},
\end{equation}
where $b$ and $\kappa$ are constants. Eq. (\ref{1}) was called a $b$-family of peakon equations.

The first equation to be distinguished within this class is the dispersionless version of the integrable Camassa-Holm (CH) equation
\begin{equation}\label{2}
u_t+2\kappa u_x-u_{xxt}+3uu_x
=2u_xu_{xx}+uu_{xxx},
\end{equation}
which is the $b=2$ case of Eq. (\ref{1}).  This equation was derived physically as a shallow water wave equation by Camassa and Holm \cite{2}. Indeed, Eq. (\ref{2}) was earlier reported by Fuchssteiner and Fokas \cite{3} as a bi-Hamiltonian generalization of the KdV equation. Eq. (\ref{2}) shares most of the important properties of an integrable system of KdV type, for example, the existence of Lax pair formalism \cite{2}, the bi-Hamiltonian structure \cite{2} and can be solved by the inverse scattering method \cite{4,5}, Darboux transformation method  \cite{6,7}  and Hirota bilinear method \cite{8} and so on \cite{9,10,19}. When $\kappa>0$, the CH equation (\ref{2}) has smooth solitary waves. It has a peculiar property that when $\kappa\rightarrow 0$ the solutions become piecewise smooth and gas corners at their crests, such solutions are weak solutions of the CH equation with $\kappa=0$ and are called peakons \cite{2,9}.  The CH equation (\ref{2}) has attracted considerable attention due to its complete integrability for all values of $\kappa$ and it has been proven that the peakons with $\kappa=0$ and all smooth solitary waves for this equation (\ref{2}) are orbitally stable \cite{11,12}. However, to the best of our knowledge, there are no reports on weak solutions related to the multi-peakon wave solution of the CH equation (\ref{2}) with  $\kappa\neq 0$. One of the aim for this paper is to derive some peakon type solutions of the CH equation (\ref{2}) with $\kappa\neq 0$. We show that it is possible to construct the multi-peakon solution of the CH equation (\ref{2}) by simply superimposing the single peakon solutions and solving for the evolution of their amplitudes and the positions of their peaks. At the same time, we further discuss their peakon-type solutions and analyze particularly dynamic behavior when the two peakons collide elastically.

For $b=3$, Eq. (\ref{1}) reduces to the Degasperis-Procesi (DP) equation \cite{13,14}
\begin{equation}\label{3}
u_t+3\kappa u_x-u_{xxt}+4uu_x
=3u_xu_{xx}+uu_{xxx}.
\end{equation}
This equation can be also be considered as a model for shallow water dynamics and be found to be completely integrable. Similar to the CH equation, the DP equation possesses the Lax pair, bi-Hamiltonian structure and also admits peakon dynamics.

Motivated by the Ref. \cite{15,16,17}, our main purpose in this paper is to investigate multi-peakon solutions of the $b$-family equations (\ref{1}) with arbitrary $b$. This paper is organized as follows. In section 2, we look for multi-peakon type solutions to the $b$-family of peakon equations (\ref{1}) arise form the solutions of the systems of ODEs. In particular, the dynamic behaviors of the two-peakon type solutions of the two cases $b=1$ and $b \neq 1$ are discussed in detail. Moreover, the dynamical systems for $N$ peakon solutions of the $b$-family of peakon equations are given. Thus, some conclusions can be concluded in section 3.

\section{Multi-peakon solutions to the $b$-family of peakon equations}

The $b$-family of peakon equations (\ref{1}) can be rewritten as
\begin{equation}\label{4}
m_{t}+b mu_{x}+um_{x}=0,\,\,\,m=u-u_{xx}+\kappa.
\end{equation}
In the next, we shall derive multi-peakon type solutions to the $b$-family of peakon equations (\ref{1}).

\subsection{One-peakon solution to the $b$-family of peakon equations}
 Let us suppose that one-peakon solution of the $b$-family of peakon equations is of
 the following form
\begin{equation}\label{5}
u(x,t)=p_{1}(t){\rm e}^{-\left|x-q_{1}(t)\right|}+r_{1}(t),
\end{equation}
where $p_{1}(t)$, $q_{1}(t)$ and $r_1(t)$ are functions of $t$
needed to be determined. In this case, it is obvious to see that function $u(x,t)$ do not has the first order derivative at the point $x=q_1(t)$, but we can obtain their derivatives $u_x$, $m$, $m_x$ and $m_t$ in the weak sense as follows
\begin{equation}\label{6}
u_{x}=-p_{1}sgn\left(x-q_{1}\right){\rm
e}^{-\left|x-q_{1}\right|},\,\,\,m=2p_{1}\delta\left(x-q_{1}\right),
\end{equation}
\begin{equation}\label{7}
m_x=2p_{1}\delta^\prime\left(x-q_{1}\right),\,\,\,\,m_t=2p_{1t}\delta\left(x-q_{1}\right)-2p_1q_{1t}\delta^\prime\left(x-q_{1}\right),
\end{equation}
where $\delta\left(x-q_{1}\right)$ denotes delta distribution function.

Substituting (\ref{5})-(\ref{7}) into Eq. (\ref{4}) and
integrating in the distribution sense, we can readily get
\begin{equation}\label{8}
p_{1t}=0,\,\,\,q_{1t}=p_1-\kappa,\,\,\,r_{1t}=0,\,\,\,r_1+\kappa=0.
\end{equation}
From (\ref{8}), it is easy to see that we may have
\begin{equation}\label{9}
p_{1}=c,\,\,\,q_{1}=(c-\kappa)t+c_0,\,\,\,r_1=-\kappa,
\end{equation}
where $c$ and $c_{0}$ are two arbitrary integration constants. Substituting
(\ref{9}) into Eq. (\ref{5}), we obtain a one-peakon
solution of the $b$-family of peakon equations (\ref{1}) as follows
\begin{equation}\label{10}
u=c\, {\rm
e}^{-\left|x-\left(c-\kappa\right)t-c_0\right|}-\kappa.
\end{equation}
For arbitrary value of $\kappa$, the unified one-peakon solution (\ref{10}) is similar to the result of the Ref. \cite{18,21}.

\subsection{Two-peakon dynamical system to the $b$-family of peakon equations}
We assume that the $b$-family of peakon equations (\ref{1}) admits two-peakon soliton as
follows
\begin{equation}\label{11}
u=p_{1}{\rm e}^{-\left|x-q_{1}\right|}+p_{2}{\rm
e}^{-\left|x-q_{2}\right|}-\kappa,
\end{equation}
where $p_{1}$ , $p_{2}$ , $q_{1}$  and $q_{2}$
are functions of $t$ needed to be determined. By a direct
calculation in the distribution sense, we have
\begin{equation}\label{12}
u_{x}=-p_{1}sgn\left(x-q_{1}\right) {\rm e}^{-\left|x-q_{1}\right|}-p_{2}sgn\left(x-q_{2}\right) {\rm
e}^{-\left|x-q_{2}\right|},
\end{equation}
\begin{equation}\label{13}
m=2p_{1}\delta
\left(x-q_{1}\right)+2p_{2}\delta\left(x-q_{2}\right).
\end{equation}
\begin{equation}\label{14}
m_x=2p_{1}\delta^\prime
\left(x-q_{1}\right)+2p_{2}\delta^\prime\left(x-q_{2}\right),
\end{equation}
\begin{equation}\label{15}
m_t=2p_{1t}\delta\left(x-q_{1}\right)-2p_{1}q_{1t}\delta^\prime
\left(x-q_{1}\right)+2p_{2t}\delta\left(x-q_{2}\right)-2p_{2t}q_{2t}\delta^\prime\left(x-q_{2}\right).
\end{equation}

Substituting (\ref{11})-(\ref{15}) into (\ref{4}) and integrating through test functions yield the following ODE dynamical system
\begin{equation}\label{16}
p_{1t}=(b-1)p_1p_2sgn\left(q_{1}-q_{2}\right){\rm e}^{-\left|q_{1}-q_{2}\right|},
\end{equation}
\begin{equation}\label{17}
p_{2t}=(b-1)p_1p_2sgn\left(q_{2}-q_{1}\right){\rm e}^{-\left|q_{2}-q_{1}\right|},
\end{equation}
\begin{equation}\label{18}
q_{1t}=p_1+p_2{\rm e}^{-\left|q_{1}-q_{2}\right|}-\kappa,
\end{equation}
\begin{equation}\label{19}
q_{2t}=p_1{\rm e}^{-\left|q_{2}-q_{1}\right|}+p_2-\kappa.
\end{equation}

We shall discuss all possible values of $b$ in the following cases.

\textbf{Case 1} $b= 1$.

Take $b= 1$, Eqs. (\ref{16})-(\ref{19}) become the following equations
\begin{equation}\label{20}
p_{1t}=p_{2t}=0,
\end{equation}
\begin{equation}\label{21}
q_{1t}=p_1+p_2{\rm e}^{-\left|q_{1}-q_{2}\right|}-\kappa,
\end{equation}
\begin{equation}\label{22}
q_{2t}=p_1{\rm e}^{-\left|q_{2}-q_{1}\right|}+p_2-\kappa.
\end{equation}

Solving Eqs. (\ref{20})-(\ref{22}), we have
\begin{equation}\label{23}
p_{1}=c_1,\,\,\,p_{2}=c_2,
\end{equation}
\begin{equation}\label{24}
q_{1}=(c_1-\kappa)t-\dfrac{c_2}{c_1-c_2}\ln(1+{\rm e}^{-(c_1-c_2)t-c_0})+C_1,
\end{equation}
\begin{equation}\label{25}
q_{2}=(c_2-\kappa)t-\dfrac{c_1}{c_1-c_2}\ln(1+{\rm e}^{-(c_1-c_2)t-c_0})+C_2,
\end{equation}
where $c_0$, $c_1$, $c_2$, $C_1$ and $C_2$ are integration constants. Substituting (\ref{23})-(\ref{25}) into (\ref{11}), we get a two-peakon solution
\begin{equation}\label{26}
u=c_{1}{\rm e}^{-\left|x-q_{1}\right|}+c_{2}{\rm
e}^{-\left|x-q_{2}\right|}-\kappa,
\end{equation}
where $q_1$ and $q_2$ are given in (\ref{24})-(\ref{25}). The dynamic behaviors of the two-peakon wave (\ref{26}) are shown in Figure 1. Figure 1 is shown two-peakon waves collide inelastically.

\textbf{Case 2} $b\neq 1$.

According to Eqs. (\ref{16})-(\ref{17}), we have
\begin{equation}\label{27}
(p_1+p_2)_t=0.
\end{equation}
By the above equation, it is easy to see that there are the
following relations
\begin{equation}\label{28}
p_{1}=p,\,\,\,p_{2}=-p+\Gamma,
\end{equation}
where $p=p(t)$ is a function of $t$, and  $\Gamma$ is a integration constant.

Without loss of generality, we take $Q=q_{1}-q_{2}>0$ and combine Eqs. (\ref{18})-(\ref{19}), we are able to get
\begin{equation}\label{29}
Q_{t}=\left(2p-\Gamma\right)\left(1-{\rm e}^{-Q}\right).
\end{equation}
From Eqs. (\ref{16})-(\ref{17}) and (\ref{28}), we have
\begin{equation}\label{30}
Q=-\ln\left(-\dfrac{p_t}{(b-1)p(p-\Gamma)}\right).
\end{equation}

Combining (\ref{29}) and (\ref{30}) leads to
\begin{equation}\label{31}
-(b-1)p(p-\Gamma)p_{tt}=(b-1)p(p-\Gamma)(2p-\Gamma)p_t+(b-2)(2p-\Gamma)p^2_t.
\end{equation}

For Eq. (\ref{31}), we shall give the following subcases $(b=2)$ and $(b\neq 1,\,\,\Gamma\neq 0)$.

\textbf{Subcase 2.1} $b=2$.

Let choosing $b=2$, we have
\begin{equation}\label{32}
p_{t}+p^2-\Gamma p=d,
\end{equation}
where $d$ is a integrate constant. For get the solutions of the Eq.
(\ref{32}), we shall discuss the following three cases.

$\mathbf{(i)}$ For $d>-\dfrac{\Gamma^{2}}{4}$, Eq.
(\ref{32}) leads to
\begin{equation}\label{33}
p=\dfrac{\Gamma}{2}+\Delta\coth\Delta(t-c_0),\,\,\,\Delta=\sqrt{d+\dfrac{1}{4}\Gamma^2}.
\end{equation}

Substituting (\ref{30}) and (\ref{33}) into (\ref{18})-(\ref{19}), we obtain
\begin{equation}\label{34}
q_1=\left(\dfrac{\Gamma}{2}-\kappa\right)t+\ln\left|\dfrac{\Gamma}{2}\sinh\Delta(t-c_0)+\Delta\cosh\Delta(t-c_0)\right|+C_1,
\end{equation}
\begin{equation}\label{35}
q_2=\left(\dfrac{\Gamma}{2}-\kappa\right)t-\ln\left|-\dfrac{\Gamma}{2}\sinh\Delta(t-c_0)+\Delta\cosh\Delta(t-c_0)\right|+C_2,
\end{equation}
where $\Delta=\sqrt{d+\dfrac{1}{4}\Gamma^2}$ and $c_0$, $C_1$, $C_2$ are integrable constants.

Therefore, we have the first type two-peakon solutions of the CH equation
(\ref{2}) as
\begin{equation}\label{36}
u=\left[\dfrac{\Gamma}{2}+\Delta\coth\Delta(t-c_0)\right]{\rm e}^{-|x-q_1|}+\left[\dfrac{\Gamma}{2}-\Delta\coth\Delta(t-c_0)\right]{\rm e}^{-|x-q_2|},
\end{equation}
where $q_1$ and $q_2$ are given in (\ref{34})-(\ref{35}). In Figure 2, we have described the interaction processes of the first type two-peakon solutions (\ref{36}).  The head on elastic collision of two-peakon waves (\ref{36}) is illustrated in Figure 2. This interaction is very similar to that of two solitons of the KdV equation.

$\mathbf{(ii)}$ For $d<-\dfrac{\Gamma^{2}}{4}$, Eq. (\ref{32}) leads to

\begin{equation}\label{37}
p=\dfrac{\Gamma}{2}+\Omega\tan\Omega(t-c_0),\,\,\,\Omega=\sqrt{-d-\dfrac{1}{4}\Gamma^2}.
\end{equation}

Substituting (\ref{30}) and (\ref{37}) into (\ref{18})-(\ref{19}), we obtain
\begin{equation}\label{38}
q_1=\left(\dfrac{\Gamma}{2}-\kappa\right)t-2\ln|\cos\Omega(t-c_0)|+\ln\left|\dfrac{\Gamma}{2}\cos\Omega(t-c_0)+\Omega\sin\Omega(t-c_0)\right|+C_1,
\end{equation}
\begin{equation}\label{39}
q_2=\left(\dfrac{\Gamma}{2}-\kappa\right)t+2\ln|\cos\Omega(t-c_0)|-\ln\left|-\dfrac{\Gamma}{2}\cos\Omega(t-c_0)+\Omega\sin\Delta(t-c_0)\right|+C_2,
\end{equation}
where $\Omega=\sqrt{-d-\dfrac{1}{4}\Gamma^2}$ and $c_0$, $C_1$, $C_2$ are integrable constants.

So we have second type two-peakon solutions for the CH equation (\ref{2})
\begin{equation}\label{40}
u=\left[\dfrac{\Gamma}{2}+\Omega\tan\Omega(t-c_0)\right]{\rm e}^{-|x-q_1|}+
\left[\dfrac{\Gamma}{2}-\Omega\tan\Omega(t-c_0)\right]{\rm e}^{-|x-q_2|},
\end{equation}
where $q_1$ and $q_2$ are given in (\ref{38})-(\ref{39}). In Figure 3, the period structure of the second type two-peakon
solutions (\ref{40}) is described. It can be seen that this solution (\ref{40}) has singularities.

$\mathbf{(iii)}$ For $d=-\dfrac{\Gamma^{2}}{4}$, Eq. (\ref{32}) leads to
\begin{equation}\label{41}
p=\dfrac{\Gamma}{2}+\dfrac{1}{t-c_0},
\end{equation}
where $c_0$ is an integrable constant.
Substituting (\ref{30}) and (\ref{41}) into (\ref{18})-(\ref{19}), we obtain
\begin{equation}\label{42}
q_1=\left(\dfrac{\Gamma}{2}-\kappa\right)t+\ln\left|\dfrac{\Gamma}{2}(t-c_0)+1\right|+C_1,
\end{equation}
\begin{equation}\label{43}
q_2=\left(\dfrac{\Gamma}{2}-\kappa\right)t-\ln\left|-\dfrac{\Gamma}{2}(t-c_0)+1\right|+C_2,
\end{equation}
where $c_0$, $C_1$ and $C_2$ are integrable constants.

So we have third type two-peakon solutions for the CH equation (\ref{2})
\begin{equation}\label{44}
u=\left[\dfrac{\Gamma}{2}+\dfrac{1}{t-c_0}\right]{\rm e}^{-|x-q_1|}+\left[\dfrac{\Gamma}{2}-\dfrac{1}{t-c_0}\right]{\rm e}^{-|x-q_2|},
\end{equation}
where $q_1$ and $q_2$ are given in (\ref{42})-(\ref{43}). Figure 4 show the profiles of the third type two-peakon
solutions (\ref{44}).

\textbf{Subcase 2.2} $b\neq 1$, $\Gamma\neq 0$.

In this case, we obtain a particular solutions of Eq. (\ref{31}) as

\begin{equation}\label{145}
p=\dfrac{A \Gamma}{A+B {\rm e}^{-(b-1)\Gamma t}},
\end{equation}
where $A$ and $B$ are integrable constants. Substituting (\ref{145}) into (\ref{28}) and (\ref{18})-(\ref{19}), we get
\begin{equation}\label{146}
p_1=\dfrac{A \Gamma}{A+B {\rm e}^{-(b-1)\Gamma t}},\,\,p_2=-\dfrac{A \Gamma}{A+B {\rm e}^{-(b-1)\Gamma t}}+\Gamma,
\end{equation}
\begin{equation}\label{147}
q_1=(\Gamma-\kappa)t+C_1,\,\,\,q_2=(\Gamma-\kappa)t+C_2,
\end{equation}
where $C_1$ and $C_2$ are integrable constants. Substituting (\ref{146})-(\ref{147}) into (\ref{11}), we have a solution of the $b$-family of peakon equations
\begin{equation}\label{148}
u=\dfrac{A \Gamma}{A+B {\rm e}^{-(b-1)\Gamma t}}\, {\rm e}^{-|x-q_1|}+\left[-\dfrac{A \Gamma}{A+B {\rm e}^{-(b-1)\Gamma t}}+\Gamma\right]{\rm e}^{-|x-q_2|},
\end{equation}
where $q_1$ and $q_2$ are given in (\ref{147}).  Figure 5 plot the solitoff structure of the $b$-family of peakon equations for six different values of $b$.

\subsection{$N$-peakon solution to the $b$-family of peakon equations}

In general, we suppose an $N$-peakon solution of the $b$-family of peakon equations (\ref{1}) has the following form
\begin{equation}\label{45}
u=\sum^N_{j=1}p_j(t){\rm e}^{-|x-q_j(t)|}-\kappa.
\end{equation}

Substituting (\ref{45}) into the $b$-family of peakon equations (\ref{1}) and integrating through test functions, we obtain the $N$-peakon dynamical system as follows
\begin{equation}\label{46}
p_{jt}=(b-1)p_j\sum^N_{k=1}p_k sgn\left(q_j-q_k\right){\rm
e}^{-\left|q_{j}-q_{k}\right|},
\end{equation}
\begin{equation}\label{47}
q_{jt}=\sum^N_{k=1}p_k {\rm e}^{-\left|q_{j}-q_{k}\right|}-\kappa,\,\,\,(j=1,2,\ldots,N).
\end{equation}
Thus, $N$-peakon solutions of the $b$-family of peakon equations with $\kappa\neq 0$  are obtained by simply superimposing the single peakon solutions and solving for the evoluting of  their amplitudes $p_j$ and the positions of their peakons $q_j$.

\section{Conclusions}
The $b$-family peakon equations (\ref{1}) are generalizing nonlinear model. Remarkably, every member (free $b$ and $\kappa$) of the $b$-family equations (\ref{1}) admits unified one-peakon solutions as form of (\ref{10}). In particular, the $b$-family peakon equations (\ref{1}) include the two special integrable models, namely the Camassa-Holm and Degasperis-Procesi equations ($b=2$ and $3$).

In this paper, the multi-peakon solutions of the $b$-family equations (\ref{1}) with $b=1$ and $b\neq1$ are discussed in detail. The results in this paper are wider than those already known  \cite{2}. At the same time, the dynamic behaviors of obtained solutions are illustrated through some figures. Taking into account the obtained results, we believe that Eqs. (\ref{46})-(\ref{47}) relating to the $b$-family equations (\ref{1}) deserves further investigation, such as in the cases of $N\geq 3$.

\section*{Acknowledgments}
This work is supported by the National Natural Science Foundation of
China under (Grant No 11261037), the Natural Science Foundation of
Inner Mongolia Autonomous Region under (Grant No 2014MS0111), the
Caoyuan Yingcai Program of Inner Mongolia Autonomous Region under
(Grant No CYYC2011050), the Program for Young Talents of Science and
Technology in Universities of Inner Mongolia Autonomous Region under
(Grant No NJYT14A04).

\begin{center}
\resizebox{3.2in}{!}{\includegraphics{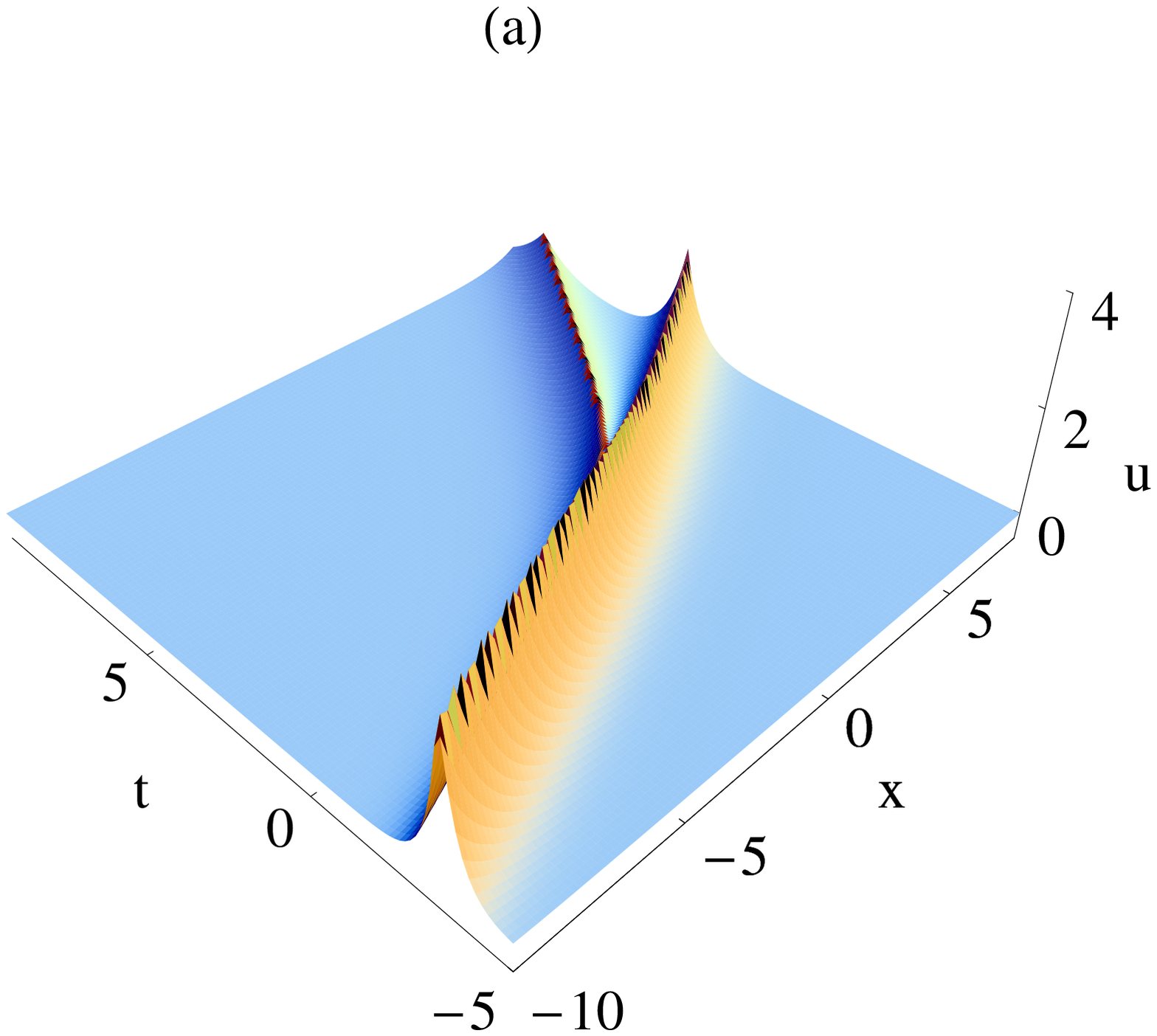}}\,\,\,
\resizebox{2.3in}{!}{\includegraphics{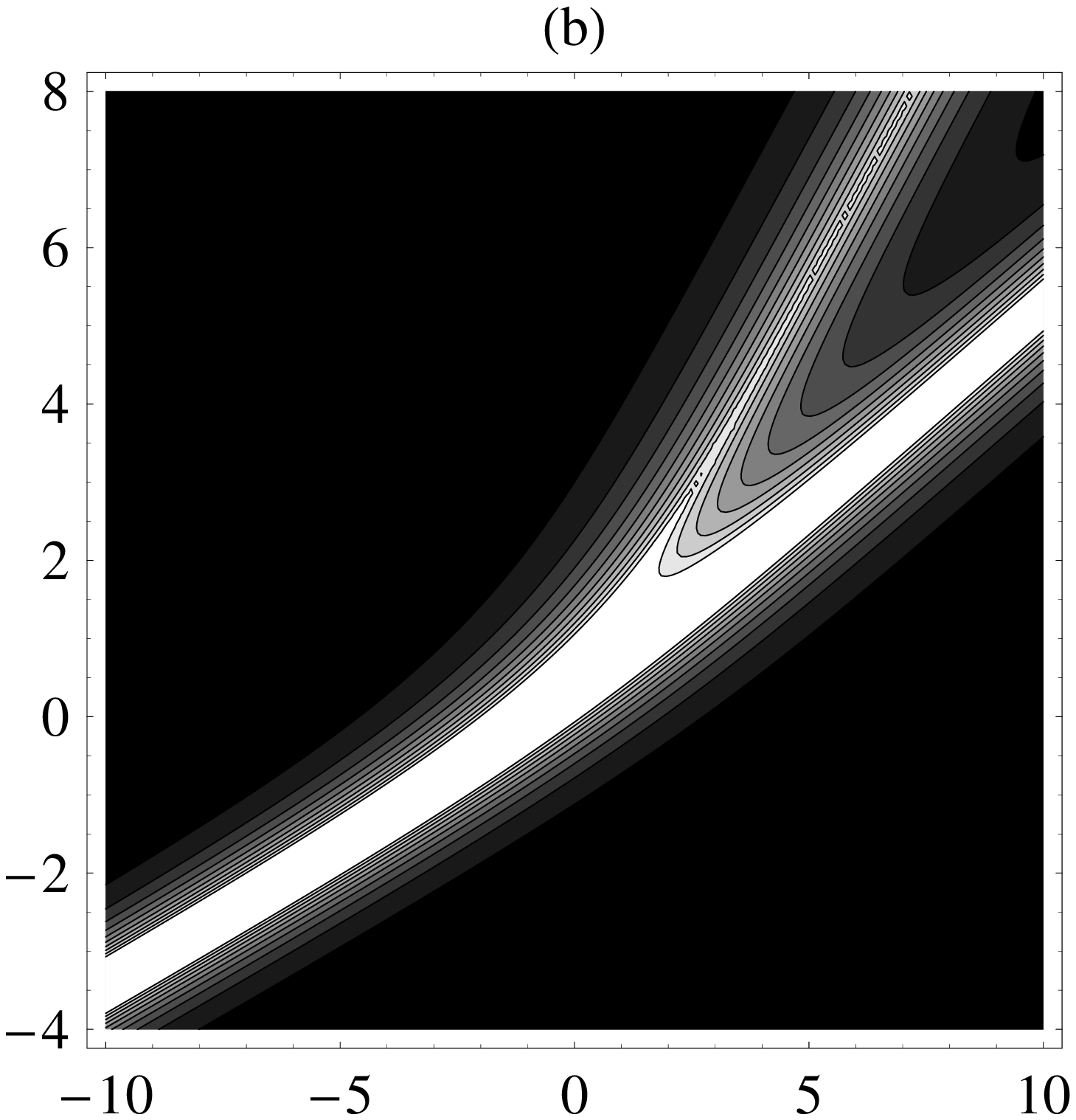}}

\end{center}
\centerline{\small{Figure 1.\,\, (a) 3D graphs of the two-peakon solutions defined by (\ref{26}) with $c_0=0$, $c_1=2$, $c_2=1$, }}
\centerline{\small{$C_1=C_2=0$, $\kappa=\frac{1}{10}$. (b) Contour plot of the two-peakon solutions defined by (\ref{26}).}}

\begin{center}
\resizebox{3.2in}{!}{\includegraphics{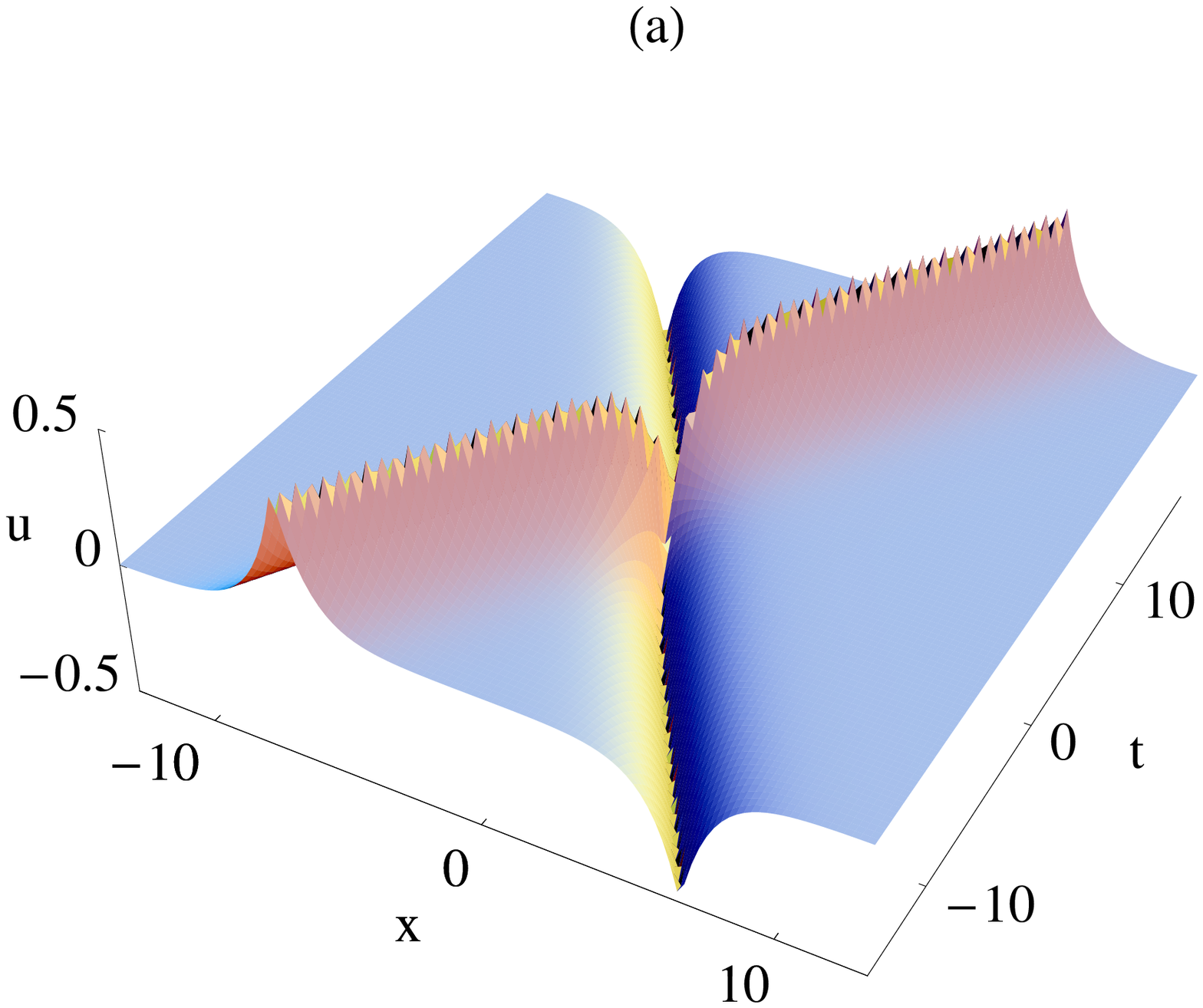}}\,\,\,
\resizebox{2.3in}{!}{\includegraphics{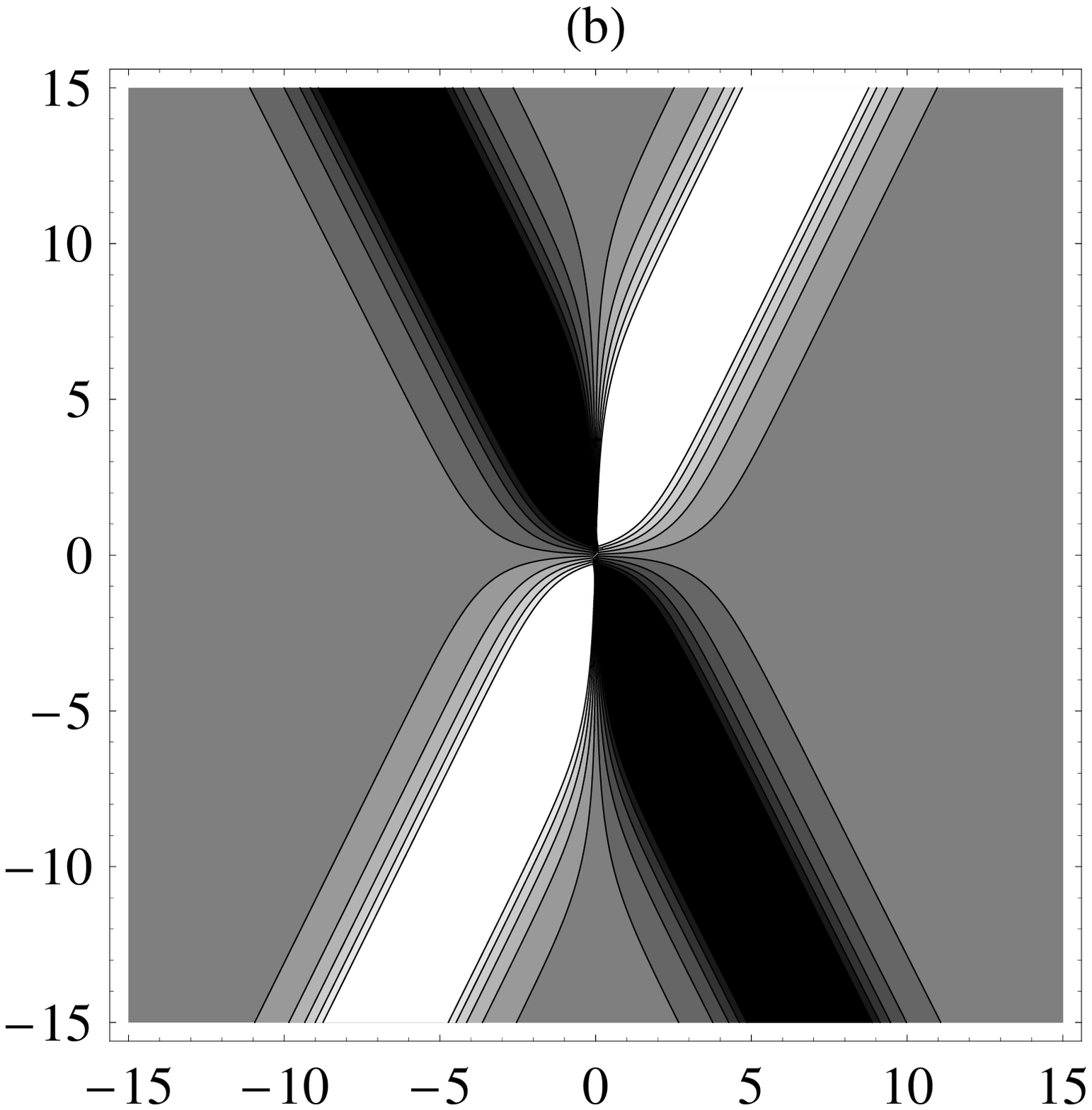}}
\end{center}
\centerline{\small{Figure 2.\,\, (a) 3D graphs of the first type two-peakon solutions defined by (\ref{36}) with $c=\frac{1}{100}$, $c_0=C_1=C_2=0$,}}
\centerline{\small{$\kappa=\frac{1}{100}$, $d=\frac{1}{4}$. (b) Contour plot of the first type two-peakon solutions defined by (\ref{36}).}}
\begin{center}
\resizebox{3.2in}{!}{\includegraphics{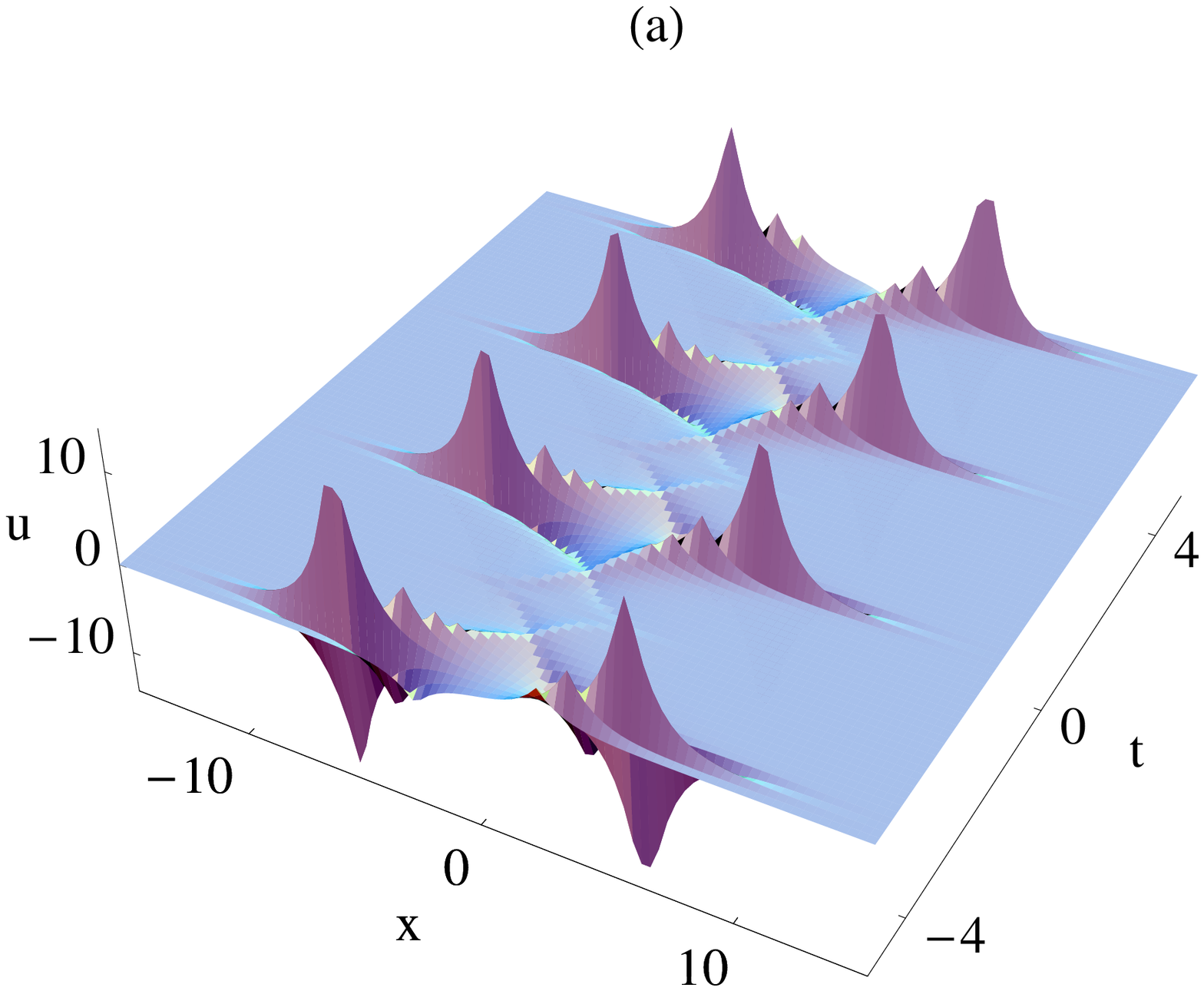}}\,\,\,
\resizebox{2.3in}{!}{\includegraphics{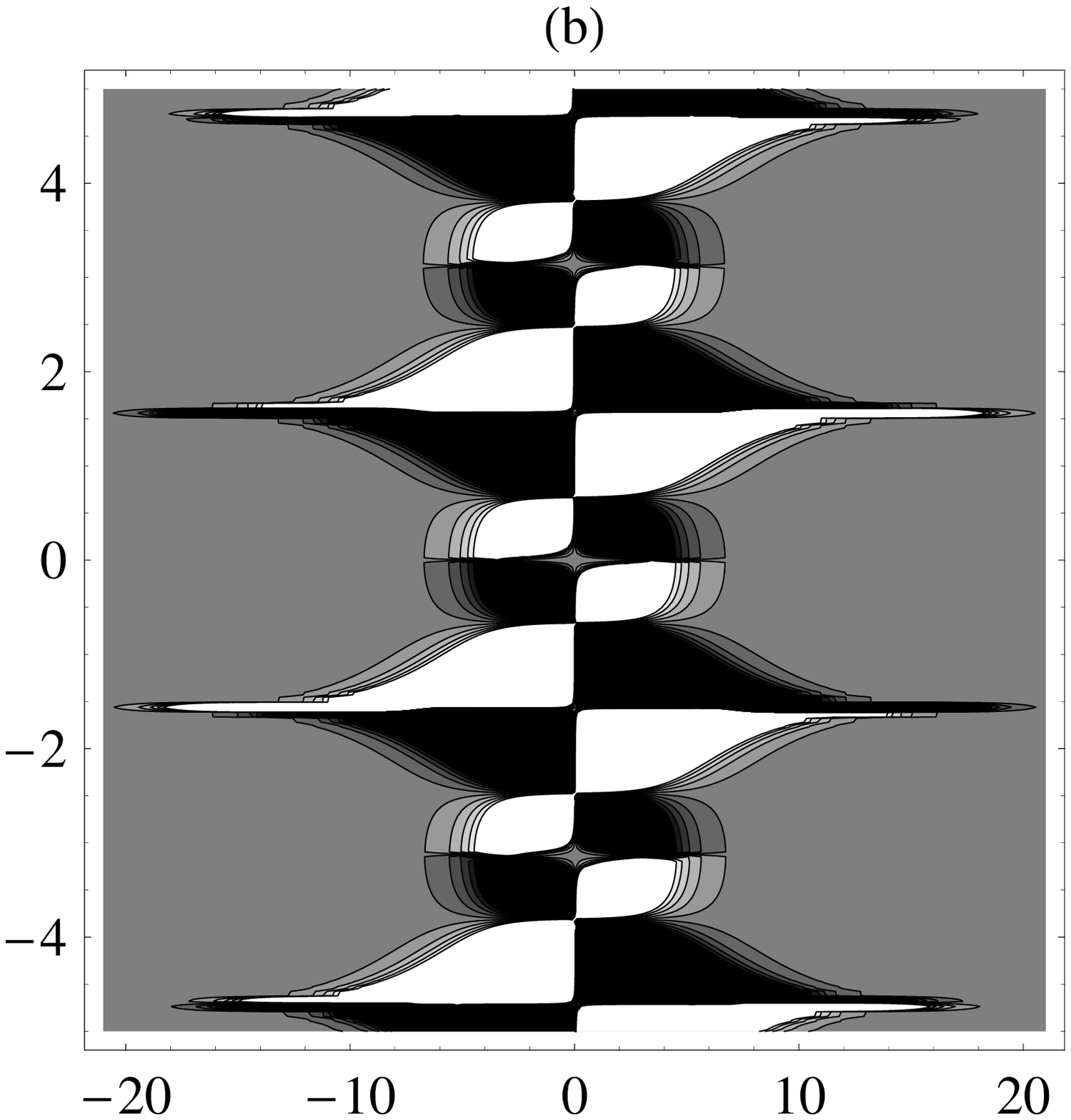}}
\end{center}
\centerline{\small{Figure 3.\,\, (a) 3D graphs of the second type two-peakon solutions defined by (\ref{40}) with $c=c_0=C_1=C_2=0$,}}
\centerline{\small{$\kappa=\frac{1}{100}$, $d=-1$. (b) Contour plot of the second type two-peakon solutions defined by (\ref{40}).}}
\begin{center}
\resizebox{3.2in}{!}{\includegraphics{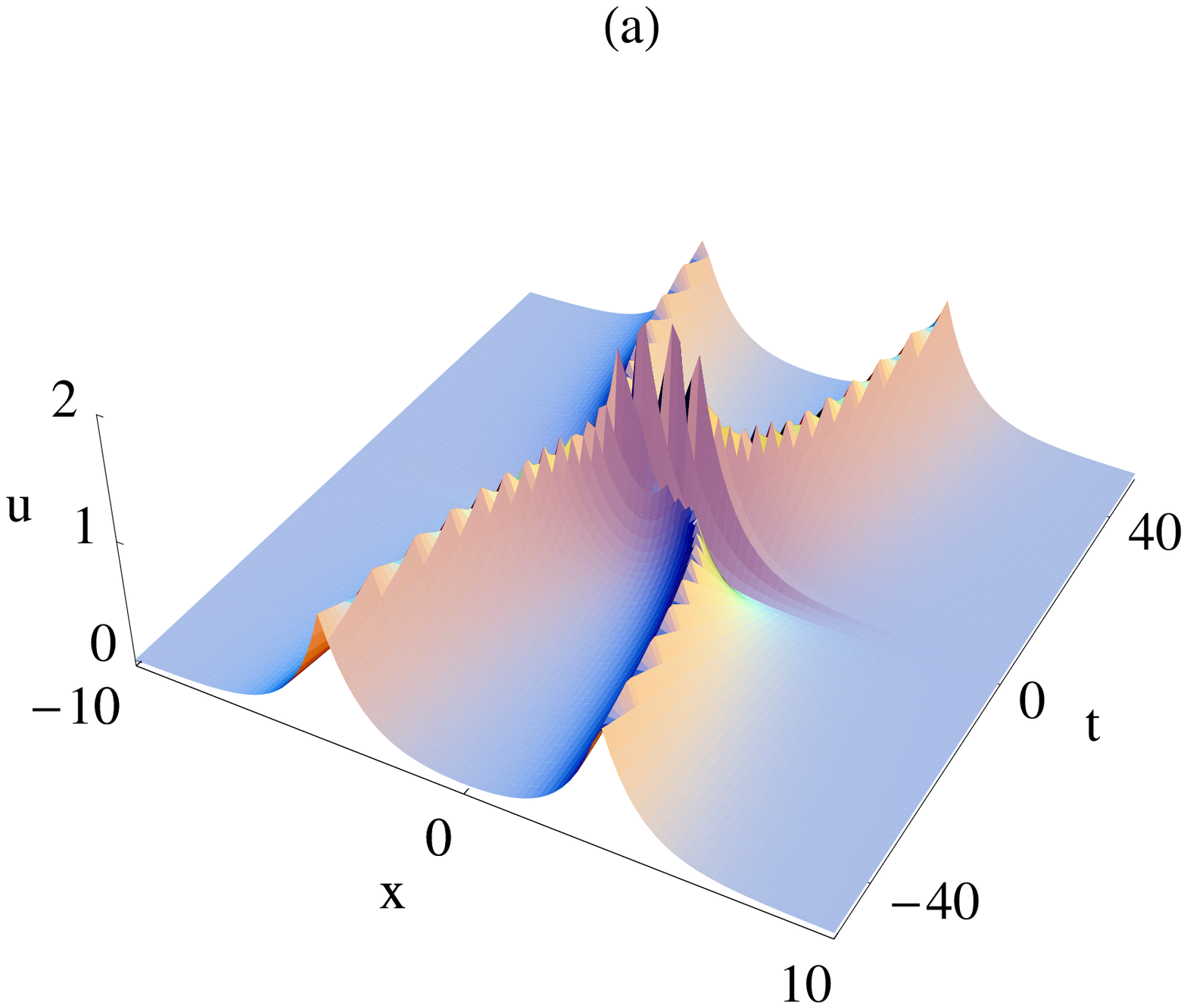}}\,\,\,
\resizebox{2.3in}{!}{\includegraphics{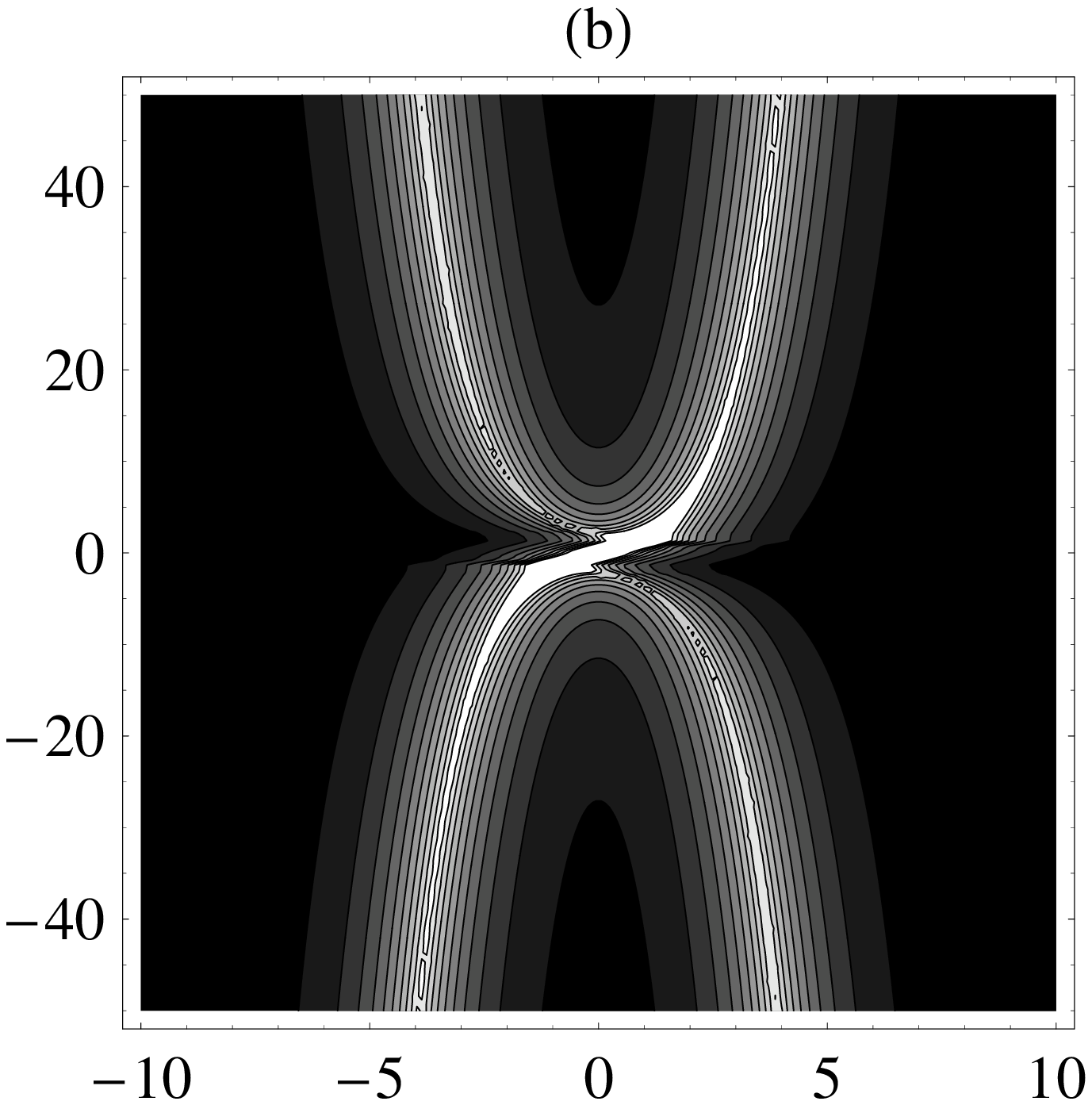}}
\end{center}
\centerline{\small{Figure 4.\,\, (a) 3D graphs of the third type two-peakon solutions defined by (\ref{44}) with $c=2$, }}
\centerline{\small{$c_0=C_1=C_2=0$, $\kappa=1$. (b) Contour plot of the third type two-peakon solutions defined by (\ref{44}).}}

\begin{center}
\resizebox{2.8in}{!}{\includegraphics{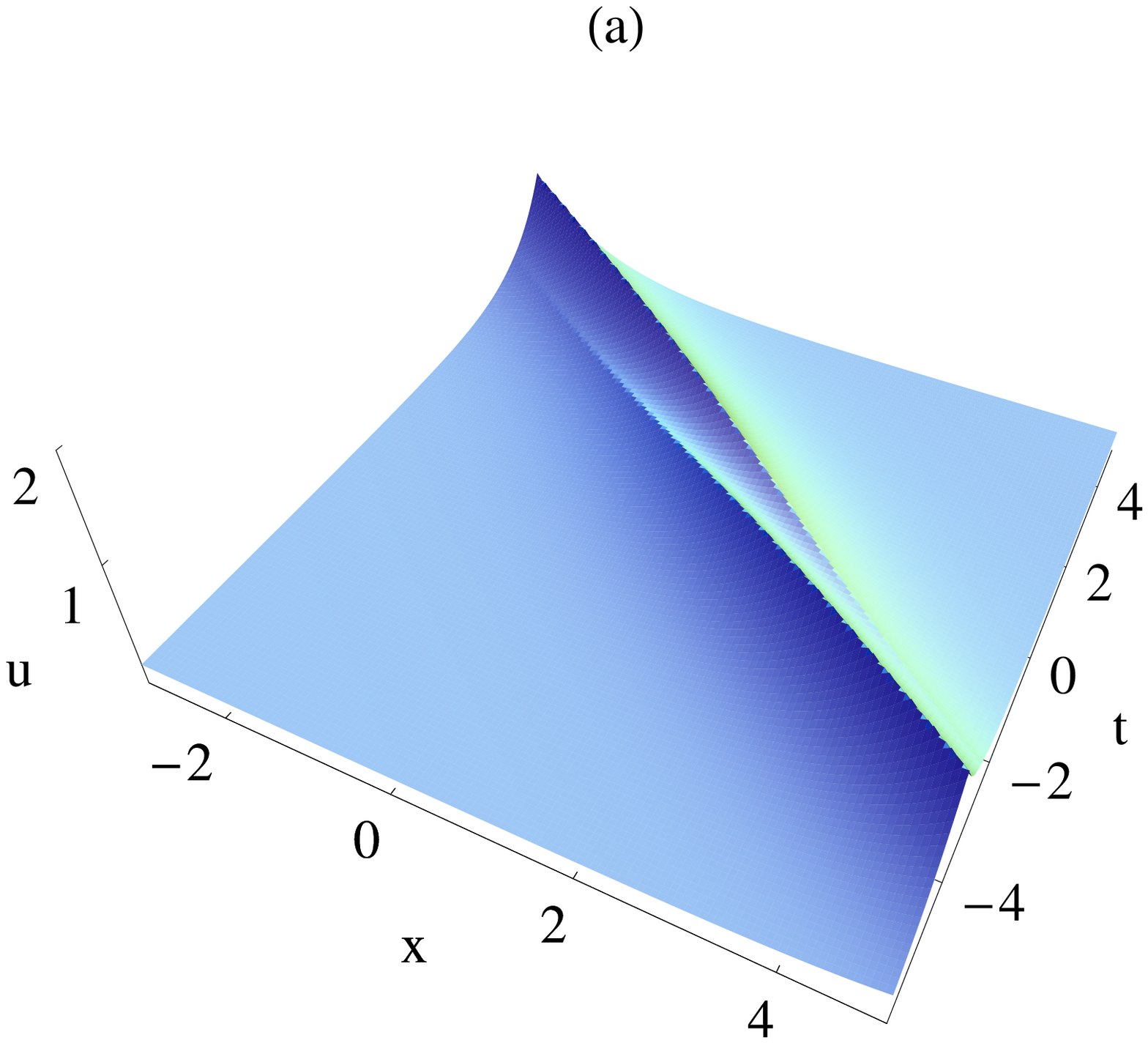}}\,\,
\resizebox{2.8in}{!}{\includegraphics{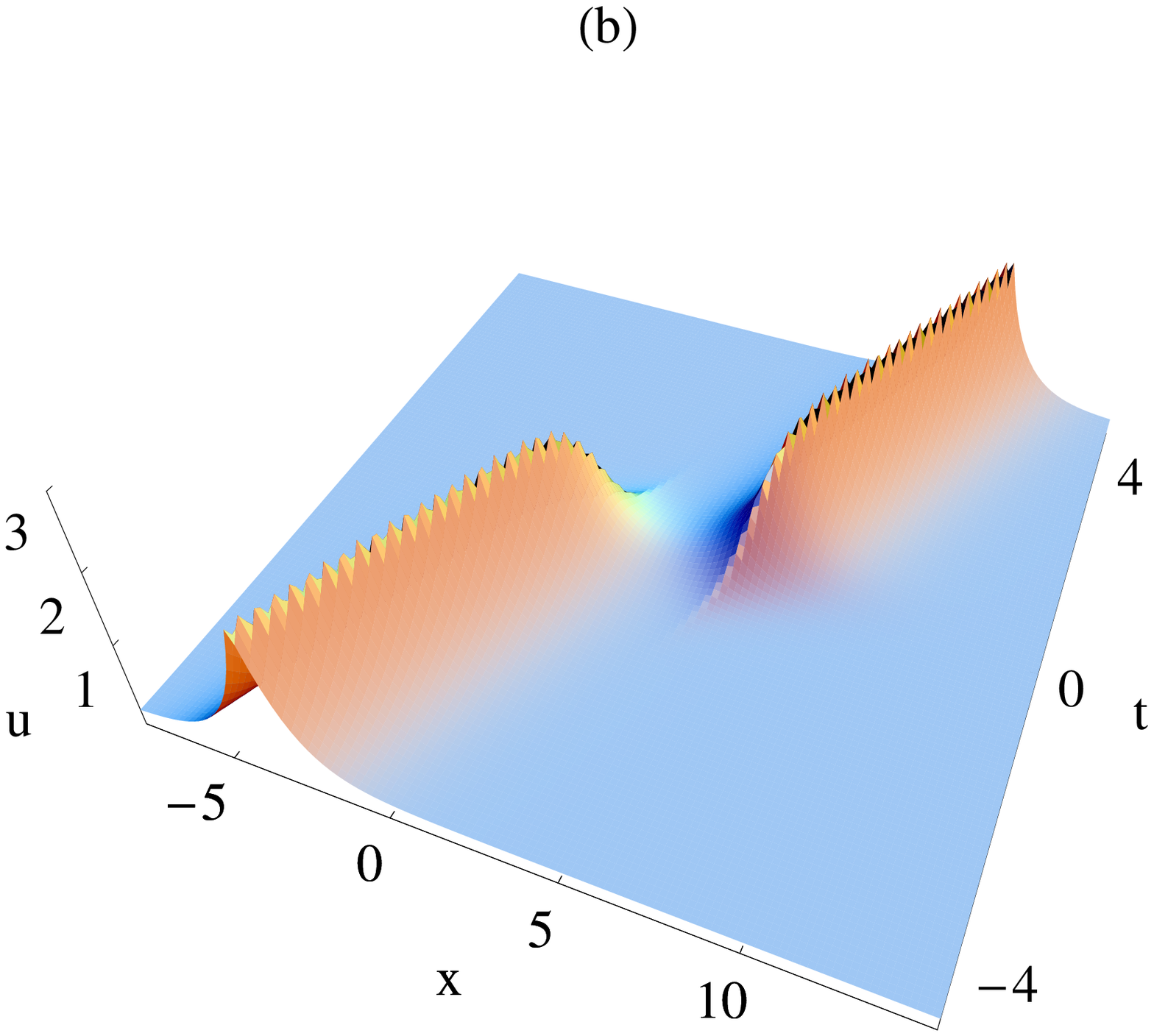}}\,\,
\resizebox{2.8in}{!}{\includegraphics{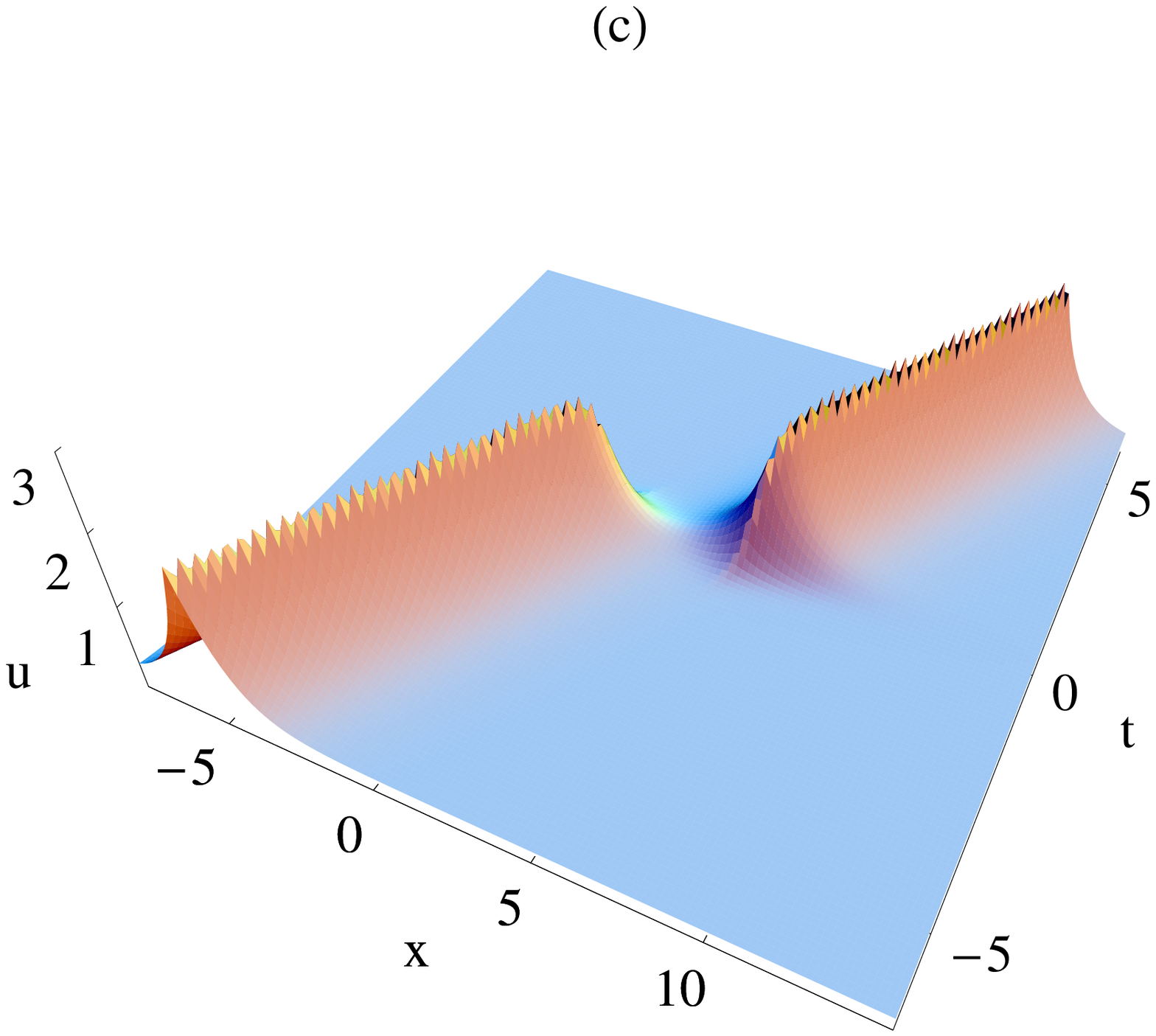}}\,\,
%\resizebox{2.8in}{!}{\includegraphics{d.eps}}\,\,
%\resizebox{2.8in}{!}{\includegraphics{e.eps}}\,\,
%\resizebox{2.8in}{!}{\includegraphics{f.eps}}\,\,
\end{center}
\centerline{\small{Figure 5.\,\, 3D graphs of the solitff structure of the solutions defined by (\ref{148}) with $C_1=6$, $C_2=1$,}}
\centerline{\small{$\Gamma=2$,\,\,$\kappa=A=B=1$, (a) $b=2$, (b) $b=3$,  (c) $b=4$.}}

%\begin{figure}[!ht]
%\centering
%\includegraphics[
%totalheight=70mm]{001.eps}
% \caption{3D graphs of the first type double
%peakon solutions defined by (\ref{14}) with
%$\Gamma=2,c=\frac{1}{2}$.}
%\end{figure}


\begin{thebibliography}{99}
\bibitem{0}Holm DD, Staley  MF, Nonlinear balances and exchange of stability in dynamics of solitons, peakons, ramp/cliffs and leftons in a 1+1 nonlinear evolutionary PDE. Phys. Lett. A 2003; 308: 437-444.
\bibitem{1}Hone  Andrew NW  and Wang JP , Prolongation algebras and Hamiltonian operators for peakon equations, Inverse Problems 2003; 19: 129-145.
\bibitem{2}Camassa R  and Holm  DD , An integrable shallow water eqution with peaked solitons, Phys. Rev. Lett. 1993; 71: 1661-1664.
\bibitem{3}Fuchssteiner B  and Fokas AS, Symplectic structures, their B\"{a}cklund transformations and hereditary symmetries. Phys. D 1981; 4: 47-66.
\bibitem{4}Constantin A, On the inverse spectral problem for the Camassa-Holm equation, J. Funct. Anal. 1998; 155: 352-363.
\bibitem{5}Constantin A, Gerdijikov  VS, Ivanov RI, Inverse scattering for the Camassa-Holm equation,  Inverse Problems 2006; 22 :2197-2207.
\bibitem{6}Li YS  and Zhang  JE, the multiple-soliton solution of the Camassa-Holm equation, Proc. R. Soc. Lond. A 2004; 460: 2617-2627.
\bibitem{7}Dai HH , Li  YS ang Su  T, Multi-soliton and multi-cuspon solutions of a Cammassa-Holm hierarchy and their interactions, J Phys. A: Theor. 2009; 42: 055203.
\bibitem{8}Parker A, On the Camassa-Holm equation and a direct method of solution. II. soliton solutions, Proc. R. Soc. Lond. A 2005; 461: 3611-3632.
\bibitem{9}Cheng  M, Liu SQ and Zhang  YJ, A two-component generalization of the Camassa-Holm equation and its solutions, Lett. Math. Phys., 2006; 75: 1-5.
\bibitem{10}Li HM ,Li  YQ and Chen Y, Reciprocal transformations of two Camassa-Holm equations, Commun. Theor. Phys. 2015; 64: 619-622.
\bibitem{19}Qu CZ and Fu  Y,On a new three-component Camassa-Holm equation with peakons, Commun. Theor. Phys.2010; 53: 223-230.
\bibitem{11}Constantin A  and Strauss  W, Stability of peakons. Commun. Pure Appl. Math. 2000; 53: 603-610.
\bibitem{12}Constantin A and Molinet L, Orbital stability of solitary waves for a shallow water equation. Physica D 2001;157: 75-89.
\bibitem{13}Degasperis A, Procesi  M, in: Degasperis A, Gaeta G (Eds.) Asymtotic integrability in symmetry and pertubation Theory, World Scientific, Singapore, 1999, PP. 23-37.
\bibitem{14}Degasperis A, Hone ANW and Holm DD, A new integrable equation with solutions, Theor. Math. Phys., 2002; 133: 146-1472.
\bibitem{15}Qiao ZJ, Xia BQ and Li JB, Integrable system with peakon, weak kink, and kink-peakon interactional solutions, arXiv:1205.2028v2.
\bibitem{16}Luo L, Xia BQ  and Cao YF, Peakon solutions to supersymmetric Camassa-Holm equation and Degasperis-Procesi equation, Commun. Theor. Phys. 2013; 59: 73-79.
\bibitem{17}Li YL, Zha QL, Multi-peakon solutions for two new coupled Camassa-Holm equations, Commun. Theor. Phys.  2016; 65: 677-683.
\bibitem{18}Liu ZR , Wang RQ and Jiang  ZJ, Peaked wave solutions of Camassa-Holm equation, Chaos, Solitons and Fractals 2004; 19 :77-92.
\bibitem{21}Guo BL, Liu ZR, Periodic cusp wave solutions and single-solitons for the $b$-equation, Chaos, Solitons and Fractals 2005; 23: 1451-1463.
\end{thebibliography}
\end {document}